# A Taxonomy of Stereotype Content in Large Language Models

**Gandalf Nicolas**[1]

**Aylin Caliskan**[2]

*This manuscript has not yet been published in a peer-reviewed outlet.*

[1] Department of Psychology, Rutgers University, New Brunswick, New Jersey 08873; gandalf.nicolas@rutgers.edu

[2] The Information School, University of Washington, Seattle, Washington, 98195

**Competing Interest Statement:** The authors do not have any competing interests.

**Abstract**

This study introduces a taxonomy of stereotype content in contemporary large language models (LLMs). We prompt ChatGPT 3.5, Llama 3, and Mixtral 8x7B, three powerful and widely used LLMs, for the characteristics associated with 87 social categories (e.g., gender, race, occupations). We identify 14 stereotype dimensions (e.g., Morality, Ability, Health, Beliefs, Emotions), accounting for ~90% of LLM stereotype associations. Warmth and Competence facets were the most frequent content, but all other dimensions were significantly prevalent. Stereotypes were more positive in LLMs (vs. humans), but there was significant variability across categories and dimensions. Finally, the taxonomy predicted the LLMs' internal evaluations of social categories (e.g., how positively/negatively the categories were represented), supporting the relevance of a multidimensional taxonomy for characterizing LLM stereotypes. Our findings suggest that high-dimensional human stereotypes are reflected in LLMs and must be considered in AI auditing and debiasing to minimize unidentified harms from reliance in low-dimensional views of bias in LLMs.

## A Taxonomy of Stereotype Content in Large Language Models

Humans create and place each other into social categories (e.g., in terms of gender, race, age, occupations) to simplify and navigate the social world, often via potentially harmful stereotypes (*1*). A stereotype is defined here and in general psychological models as a characteristic associated with a social category (e.g., through explicit beliefs, implicit associations, *2*). These stereotypes can vary in content (i.e., what they are about) and valence (i.e., how positive or negative they are), among other properties (*3*). Recent models have used text analysis to characterize the diversity of stereotypes across salient social categories in human survey data (*4*). However, stereotype content has not been systematically described for contemporary Artificial Intelligence (AI) large language models (LLMs). Effective auditing and potential debiasing solutions for AI bias require first a more comprehensive taxonomy of the various stereotypes that are associated with social categories in LLMs. Thus, the current paper uses a multimethod approach, including methods such as cluster and dictionary analyses, to introduce a taxonomy of stereotype content in three state-of-the-art LLMs.

The taxonomy includes various dimensions of content (i.e., distinct semantics groupings in terms of what the stereotypes are about, including: Sociability, Morality, Ability, Assertiveness, socioeconomic Status, political-religious Beliefs, Health, Occupation, Emotion, Deviance, Social Groups, Geographic origin, and Appearance), and we characterize how well these dimensions account for the LLMs' associations with social categories (i.e., coverage). We also describe the content dimensions' properties (direction and valence), change over responses, and correspondence with human evaluations of social categories (i.e., their predictive value).

### Stereotype Content in Human Data

The best-established stereotype dimensions are Warmth and Competence, which are evolutionarily plausible, have been found cross-culturally and over time, and are predictive of emotions and behaviors (*5*, *6*). Warmth (also called communion or the horizontal dimension) refers to attributions about a target's sociability and morality. Competence (also called agency or the vertical dimension) refers to attributions about a target's abilities and assertiveness. In other words, when evaluating others, humans prioritize understanding: is this person a friend or foe (Warmth), and can they act on their intentions (Competence)? Expanding into a more comprehensive taxonomy, the Spontaneous Stereotype Content Model (SSCM, *4*) suggested that about 14 content dimensions account for 80-95% of freely-generated stereotypes about salient social categories. These dimensions are: Sociability and Morality (facets of Warmth), Ability and Assertiveness (facets of Competence), socioeconomic Status, political-religious Beliefs (*7*, *8*), Appearance, Emotion, Occupation, Health, Deviance, Geography, Family relations, and Social Groups (e.g., "rich people are men"). Additionally, these dimensions vary in representativeness (i.e., prevalence of association), with Warmth and Competence facets being highly representative across social categories, while associations about Health and Geographic origin are less prevalent (*4*). Representativeness relates to primacy and importance of stereotype dimensions (*3*).

In addition to representativeness, many stereotype dimensions vary in direction, which refers to how "low" to "high" a dimension is, along opposite ends of a semantic differential. To illustrate, a category can be evaluated as low Morality (e.g., "immoral," "dishonest") or high Morality (e.g., "moral," "honest); low Deviance (e.g., "unremarkable," "average"), or high Deviance (e.g., "unique," "weird," *4*).

A related property is valence, that is, how positive vs. negative the stereotype is. Direction and valence often correlate. For example, being "uneducated" or "unintelligent" (low Ability) is evaluated more negatively than being "educated" or "intelligent" (high Ability). However, the two constructs are differentiable. For example, high Assertiveness can sometimes be seen negatively ("aggressiveness"), and some dimensions, such as political-religious Beliefs vary in direction from liberal/secular to conservative/religious (whether liberal or conservative is at the low or high endpoint is arbitrary; as used, this dimension effectively measures conservatism), which tend not to strongly correlate with valence on average. Additionally, valence is more general, which has strengths, such as being applicable to more content dimensions and to overall evaluations, and thus making it a strong signal in language (*9*). However, this generality also has limitations, such as having more variability across contexts (e.g., what is positive/negative may vary based on domain, *4*).

Finally, the representativeness and direction/valence of a dimension are differentiable properties of stereotypes. For example, using numerical scales, Americans evaluate nurses and doctors as being similarly high/positive Warmth and high Competence (i.e., direction/valence). However, when open-endedly describing these targets, associations for nurses are mostly *about* Warmth, while associations for doctors are mostly *about* Competence. That is, despite similar direction/valence, Warmth is more representative of stereotypes about nurses, while Competence is more representative of stereotypes about doctors (*4*).

**Stereotype Content in LLMs**

LLMs (e.g., ChatGPT, *10–12*) are generative AI models trained on vast amounts of text data which learn contextualized semantics and can generate human language in response to linguistic input. Given their training data from the internet and other text sources, LLMs reproduce many human stereotypes and biases (*13*). However, almost all the research on the topic has examined either general valence associations, a very limited number of stereotype dimensions, or simple word associations (without identifying more generalizable dimensions of meaning). For example, research shows that many social categories have negative representations in AI models paralleling human stereotypes (*14–16*), and that Warmth and Competence valence differences emerge in LLM stereotypes (*17–20*). More recent papers have looked at 3-dimensional models (e.g., Warmth/communion, Competence/status, and Beliefs, *8*, *21*). However, these are still low-dimensional models that may not account for a near-totality of stereotypical associations in LLMs. Whether more dimensions (and which) are needed to understand social representations in AI, has yet to be systematically examined.

Additionally, most research has focused on specific social categories, such as gender or race (*22*, *23*). However, understanding more generalizable patterns of stereotype content requires examining larger and more representative samples of categories (*5*). Previous research has also focused on examining text embeddings directly (*24*, *25*), the numerical representations of text that underlie the more conversational output of LLMs that users interact with through chatbots. Here, we focus on the text output directly, as these constitute the final product in most applications and have the most direct impact on the general public.

**Consequences of Stereotypical Associations**

Stereotypes may be inaccurate, over-generalized, essentializing, and self-fulfilling, among other well-documented problematics (*26*), often resulting in discrimination, conflict, and adverse health impacts for stigmatized groups (*27*, *28*). Both positive and negative stereotypes can be harmful (*29*), and their effects have been thoroughly documented. For example, stereotype content predicts outcomes such as emotional responses and interpersonal behaviors (*30*), hiring and performance evaluations (*31*), interactions across societal and organizational hierarchies (*32*, *33*), and attitudes towards AI (*34*).

These consequences may be amplified, and stereotypes reinforced, via biased AI models. LLMs have become ubiquitous in applications with real-world impact, from healthcare (*35*) to hiring (*36*). As with research and auditing, efforts to minimize harms from LLM stereotypes have so far focused on general valence, or in a few cases, on a limited number of dimensions and/or a small number of social categories (*17*, *18*). However, a more comprehensive taxonomy of the various LLM stereotype dimensions may be needed for more effective auditing and potential debiasing solutions for AI fairness.

**Current Study**

In the current study we introduce a taxonomy of stereotype content in contemporary LLMs, including identifying dimensions using both data- and theory-driven approaches, and characterizing the coverage, representativeness, direction, valence, change over responses, and predictive value of the diverse set of proposed content dimensions. We derive this taxonomy based on the models' semantic associations to several U.S. social category terms, in line with social psychological models focusing on generalizable stereotype properties that are applied across social categories. We focus on three recent and widely used LLMs: ChatGPT 3.5 Turbo (primary model), as well as Llama 3 and Mixtral 8x7B Instruct (replication models for a subset of analyses), providing convergent evidence for the taxonomy.

We prompted the LLMs to list 50 characteristics associated with social categories and coded responses into semantic dimensions using cluster and dictionary analyses. We examine how many of the LLM associations are coded into meaningful dimensions (i.e., coverage) and how frequently the different dimensions occurred in stereotypes across categories (i.e., representativeness/prevalence). Additionally, we explore changes in content representativeness across the 50 characteristics requested. We also measured how high vs. low (direction) and positive vs. negative (valence) the different dimensions are. Finally, we examined whether the full taxonomy predicts general valence evaluations of the categories (by both the LLMs and humans).

We expected to find significant overlap with the human SSCM data (*4*), where dimensions related to Warmth (Sociability, Morality) and Competence (Ability, Assertiveness) are highly representative, but not sufficient to characterize stereotype content, with additional dimensions (e.g., Emotion, Deviance) showing significant prevalence across categories. Given chatbots' safety features (*37*), we expected either similar or more positive stereotypes than human data for ChatGPT and Mixtral, and more negative stereotypes for Llama 3's base model. Finally, we hypothesized the expanded taxonomy would add significant predictive value above current valence-exclusive approaches, suggesting better capture of both internal and human social category representations in LLMs.

## Results

### Cluster Analysis

A majority of fit metrics suggested either 2 or 59 clusters. To capture meaningful diversity in content, we used a 59-cluster solution (see online repository for more information; Table 1 shows cluster examples).

As expected, this data-driven approach revealed multiple clusters related to dimensions from psychological models. A replication cluster analysis using Llama3 with both Llama3 and SBERT embeddings also showed evidence for a taxonomy composed of these dimensions (see Supplement).

**Table 1. Example clusters for dimension identification, ChatGPT.**

| **Morality** | **Sociability** | **Ability** | **Assertiveness** | **Beliefs** | **Status** | **Emotions** |
|---|---|---|---|---|---|---|
| ***Malicious*** | ***Social orientation*** | ***Analytical*** | ***Perseverance*** | ***Religion*** | ***Wealth*** | ***Negative emotion*** |
| malicious | unsympathetic | analytical thinker | perseverance | religious | opulent | dejected |
| nasty | unempathetic | skilled in problem-solving | striving for success | non-devout | luxury-oriented | despondent |
| villainous | inconsiderate | analytical thinking | high-achieving | religion | wealthy-looking | despairing |
| dangerous | ungrateful | skilled in critical thinking | strong-willed | bible-believing | rich-looking | unhappy |
| evil | uncooperative | analytical-thinking | toughness | devout | well-heeled | sorrowful |

| **Deviance** | **Appearance** | **Health** | **Geography** | **Family** | **Culture** | **Social Groups** |
|---|---|---|---|---|---|---|
| ***Unusualness*** | ***Body Properties*** | ***Low Health*** | ***Nationalities*** | ***Family*** | ***Music*** | ***Gender & Sexuality*** |
| unusual | puffy | medical care | slavic | family-nurturing | indie music | gender non-conforming |
| strange | flabby | health problem | slovenian | family-creating | music | sexual orientation |
| bizarre | chubby | wheelchair-bound | croatian | family-focused | music lover | lgbtq+ |
| weird | frumpy | intellectually disabled | belarusian | family-oriented | pop culture | genderqueer |
| odd | lumpy | unable to access healthcare | ukrainian | family-satisfying | music-loving | gender expression |

Two of the ChatGPT clusters captured linguistic regularities (e.g., words starting with "un") rather than stereotype content, and nine clusters had a mixture of content. In general, the cluster solution may be influenced by length of phrases, syntax, punctuation, and other non-content features encoded in the embeddings. These limitations are addressed by alternative methods. Specifically, the dimensions identified were largely overlapping with existing dictionaries. Thus, the cluster analysis provided a data-driven set of dimensions that align with the instruments we use next.

**Coverage**

The taxonomy dimensions, as measured through dictionaries, accounted for 93.5% of the ChatGPT, 94.3% of the Llama 3, and 88.4% of the Mixtral responses. Thus, the taxonomy characterizes the vast majority of top stereotypical associations for a large sample of salient social categories across three widely-used LLMs, in line with human studies(*4*). Unaccounted-for responses tended to be idiosyncratic, include names or other non-trait information, among other patterns. For comparison, the "big two" of Warmth and Competence accounted for only ~54% of responses in ChatGPT and Mixtral, and 63% of Llama 3 responses.

**Representativeness**

The various taxonomy dimensions differed in how representative they were of stereotypes across social categories, ChatGPT: $F(14, 9518.37) = 295.73$, $p < .001$; Llama 3: $F(14, 9705) = 357.66$, $p < .001$; Mixtral: $F(14, 10263.92) = 263.61$, $p < .001$, see Table 2 and Figure 1. Table results include descriptive information from the recent human studies with a similar design described previously (*4*), as a baseline for comparison. Across LLMs, the most representative dimensions were Ability and Assertiveness (facets of Competence), as well as Morality and Sociability (facets of Warmth). Beliefs and Status content followed, in line with their relevance in human research (*7*). Appearance and Emotions were also highly prevalent. Associations about Geographic origin (e.g., "foreign") were present, but less frequently. An "Other" category groups less frequent content related to culture, family, fortune, arts, and science.

For representativeness, we show response rates in figures and proportions in tables and main analyses (for clarity of presentation, but results are congruent across both metrics/methods).

**Table 2. Prevalence for each dimension of the taxonomy, dictionary analysis.**

| Dimension | Human baseline | Dimension | ChatGPT | Dimension | Mixtral | Dimension | Llama 3 |
|---|---|---|---|---|---|---|---|
| Ability | 0.177 | Ability | 0.209[a] | Ability | 0.177[a] | Morality | 0.247[a] |
| Morality | 0.158 | Assertiveness | 0.199[a] | Morality | 0.173[a] | Assertiveness | 0.206[b] |
| Sociability | 0.157 | Morality | 0.17[b] | Assertiveness | 0.166[a] | Sociability | 0.176[c] |
| Assertiveness | 0.142 | Sociability | 0.125[c] | Sociability | 0.123[b] | Ability | 0.168[c] |
| Status | 0.094 | Beliefs | 0.102[d] | Status | 0.093[c] | Emotions | 0.071[d] |
| Appearance | 0.08 | Status | 0.094[d] | Beliefs | 0.08[c] | Beliefs | 0.066[d] |
| Emotion | 0.076 | Appearance | 0.066[e] | Appearance | 0.076[c] | Appearance | 0.062[de] |
| Beliefs | 0.069 | Emotions | 0.05[ef] | Emotions | 0.055[d] | Status | 0.06[de] |
| Deviance | 0.038 | Health | 0.036[fg] | Occupation | 0.051[d] | Health | 0.045[ef] |
| Health | 0.033 | Occupation | 0.035[fg] | Other | 0.043[de] | Other | 0.028[fg] |
| Occupation | 0.023 | Other | 0.034[fg] | Health | 0.038[def] | Social Groups | 0.022[g] |
| Other | 0.022 | Deviance | 0.025[g] | Geography | 0.028[ef] | Deviance | 0.022[g] |
| Social Groups | 0.021 | Social Groups | 0.022[g] | Social Groups | 0.024[f] | Occupation | 0.015[g] |
| Geography | 0.015 | Geography | 0.02[g] | Deviance | 0.022[f] | Geography | 0.012[g] |
| **Correlation to human baseline** | | | .942 | | .958 | | .935 |

*Note*. Human baseline presented retrieved from the SSCM paper (*4*). Rows within a column with the same superscript are not significantly different from each other ($p > .05$). Results shown as average (across categories) proportions of responses about the dimension. The correlation to human baseline row shows the correlation between each language model and the human baseline, using average dimension prevalence (column values) as observations.

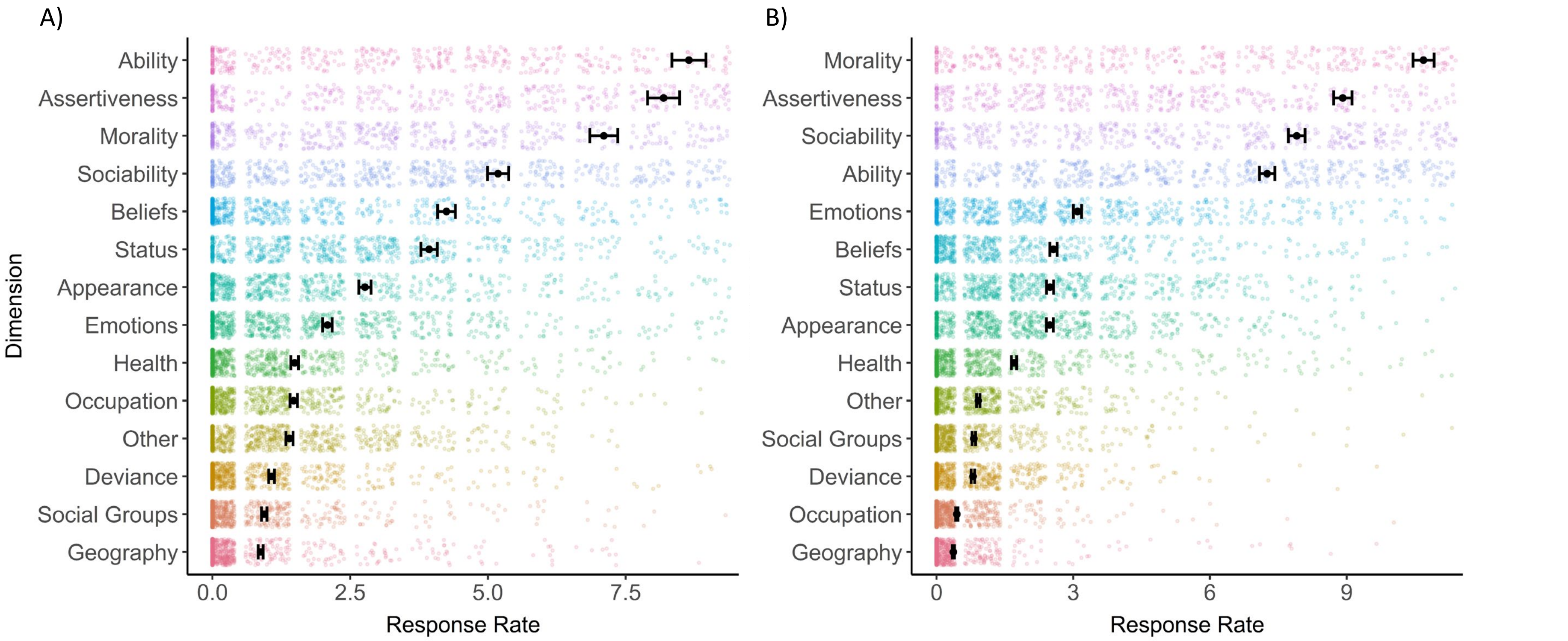

**Fig. 1. Prevalence for each dimension of the taxonomy, dictionary coding.** Response rates indicate average rates across categories, out of 50 responses. Error bars indicate Standard Errors. X-axis is truncated for presentation, see online repository for full figure. Panels: A) ChatGPT, B) Llama 3. The Mixtral figure is very similar to the ChatGPT figure and is included in the Supplement.

**Direction**

The direction of the various dimensions differed significantly, ChatGPT: $F(7, 4008.02) = 130.64$, $p < .001$; Llama 3: $F(7, 3838.52) = 107.03$, $p < .001$; Mixtral: $F(7, 4483.30) = 229.79$, $p < .001$, see Table 3 and Figure 2. Most dimensions were high-directional on average, with highest scores for Deviance, Ability, and Assertiveness, and negative direction for Health (as well as for Sociability and Morality in Llama 3).

**Table 3. Prevalence for each dimension of the taxonomy, dictionary analysis.**

| Dimension | Human baseline | Dimension | ChatGPT | Dimension | Mixtral | Dimension | Llama 3 |
|---|---|---|---|---|---|---|---|
| Assertiveness | 0.482 | Deviance | 0.588[a] | Ability | 0.666[a] | Assertiveness | 0.44[a] |
| Ability | 0.245 | Ability | 0.554[a] | Deviance | 0.568[ab] | Deviance | 0.366[ab] |
| Status | 0.169 | Assertiveness | 0.481[a] | Assertiveness | 0.519[bc] | Beliefs | 0.25[b] |
| Beliefs | -0.001 | Morality | 0.299[b] | Sociability | 0.449[c] | Ability | 0.226[b] |
| Sociability | -0.02 | Sociability | 0.275[b] | Morality | 0.447[c] | Status | 0.1[c] |
| Morality | -0.09 | Status | 0.275[b] | Status | 0.314[d] | Sociability | -0.135[d] |
| Deviance | -0.522 | Beliefs | 0.051[c] | Beliefs | 0.012[e] | Morality | -0.144[d] |
| Health | -0.747 | Health | -0.401[d] | Health | -0.499[f] | Health | -0.559[e] |
| **Correlation to human baseline** | | | .555 | | .600 | | .601 |

*Note*. Direction ranges from -1 (low) to +1 (high). For Beliefs, low indicates liberalism while high indicates conservatism. Human baseline presented retrieved from the SSCM paper (*4*). Rows within a column with the same superscript are not significantly different from each other ($p > .05$).

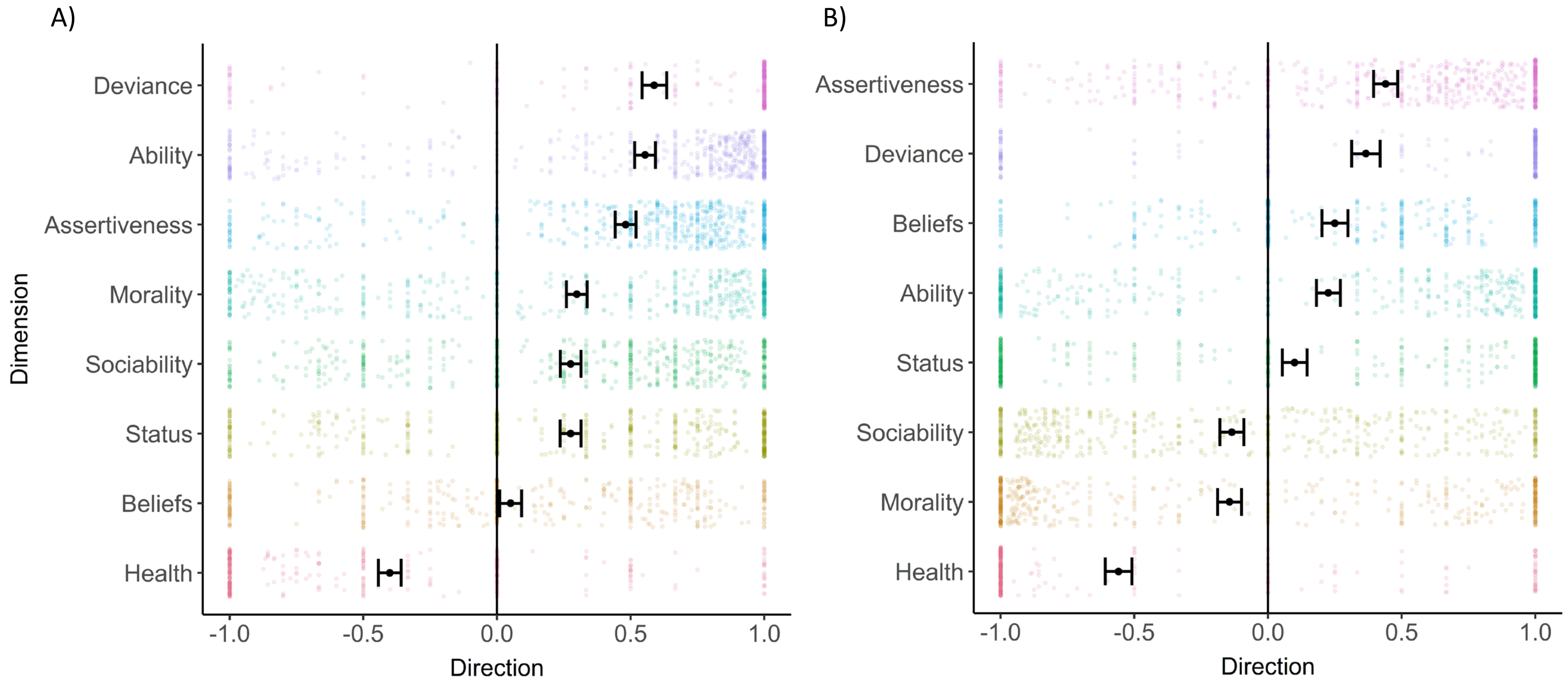

**Fig. 2. Direction for each dimension of the taxonomy.** Direction ranges from -1 (low) to +1 (high). For Beliefs, low indicates liberalism while high indicates conservatism. Error bars indicate Standard Errors. Panels: a) ChatGPT, b) Llama 3. The Mixtral figure is very similar to the ChatGPT figure and is included in the Supplement.

**Valence**

Collapsing across dimensions, the valence of the associations was positive for the chatbots with safeguards, ChatGPT: $M = .187$, $t(84.3) = 6.154$, $p < .001$; Mixtral: $M = .262$, $t(85.8) = 10.31$, $p < .001$; and neutral for the Llama 3 base model: $M = -.03$, $t(84.8) = 0.837$, $p = .405$. However, the valence of specific dimensions differed significantly, ChatGPT: $F(13, 6200.68) = 19.1$, $p < .001$; Llama 3: $F(13, 5707.56) = 11.8$, $p < .001$; Mixtral: $F(13, 7107.1) =35.6$, $p < .001$, see Table 4 and Figure 3. Valence showed more significant differences between Llama 3 and the other models, potentially as a result of its lack of chatbot finetuning, which may reduce its level of safeguards targeting valence. Llama 3 was more negative across dimensions, more in line with human data. However, the average relative valence of specific dimensions were more similar between human data and ChatGPT and Mixtral.

**Table 4. Valence for each dimension of the taxonomy.**

| Dimension | Human baseline | Dimension | ChatGPT | Dimension | Mixtral | Dimension | Llama 3 |
|---|---|---|---|---|---|---|---|
| Ability | 0.135 | Ability | 0.334[a] | Ability | 0.442[a] | Social Groups | 0.112[a] |
| Assertiveness | 0.09 | Status | 0.289[ab] | Morality | 0.417[a] | Geography | 0.09[ab] |
| Occupation | 0.074 | Morality | 0.272[abc] | Status | 0.328[b] | Ability | 0.078[a] |
| Status | 0.061 | Emotion | 0.27[abcd] | Emotion | 0.327[bc] | Status | 0.066[a] |
| Sociability | 0.04 | Appearance | 0.253[bcde] | Sociability | 0.315[bcd] | Appearance | 0.058[a] |
| Social Groups | 0.014 | Sociability | 0.216[cdef] | Appearance | 0.275[bcde] | Other | 0.041[ab] |
| Deviance | 0.001 | Assertiveness | 0.193[ef] | Assertiveness | 0.257[cde] | Deviance | 0.023[abc] |
| Morality | -0.022 | Other | 0.192[def] | Deviance | 0.251[bcdef] | Emotion | 0.021[ab] |
| Emotion | -0.055 | Beliefs | 0.192[def] | Beliefs | 0.246[def] | Assertiveness | -0.06[bcd] |
| Beliefs | -0.058 | Health | 0.175[efg] | Other | 0.237[ef] | Health | -0.07[bcd] |
| Geography | -0.072 | Deviance | 0.163[fgh] | Health | 0.219[ef] | Occupation | -0.086[bcd] |
| Other | -0.08 | Occupation | 0.135[fgh] | Social Groups | 0.166[fg] | Morality | -0.089[cd] |
| Appearance | -0.084 | Geography | 0.105[gh] | Occupation | 0.11[g] | Beliefs | -0.105[d] |
| Health | -0.307 | Social Groups | 0.082[gh] | Geography | 0.101[g] | Sociability | -0.122[d] |
| **Correlation to human baseline** | | | .230 | | .267 | | .108 |

*Note*. Valence ranges from -1 (negative) to +1 (positive). Human baseline presented retrieved from the SSCM paper(*4*). Rows within a column with the same superscript are not significantly different from each other ($p > .05$).

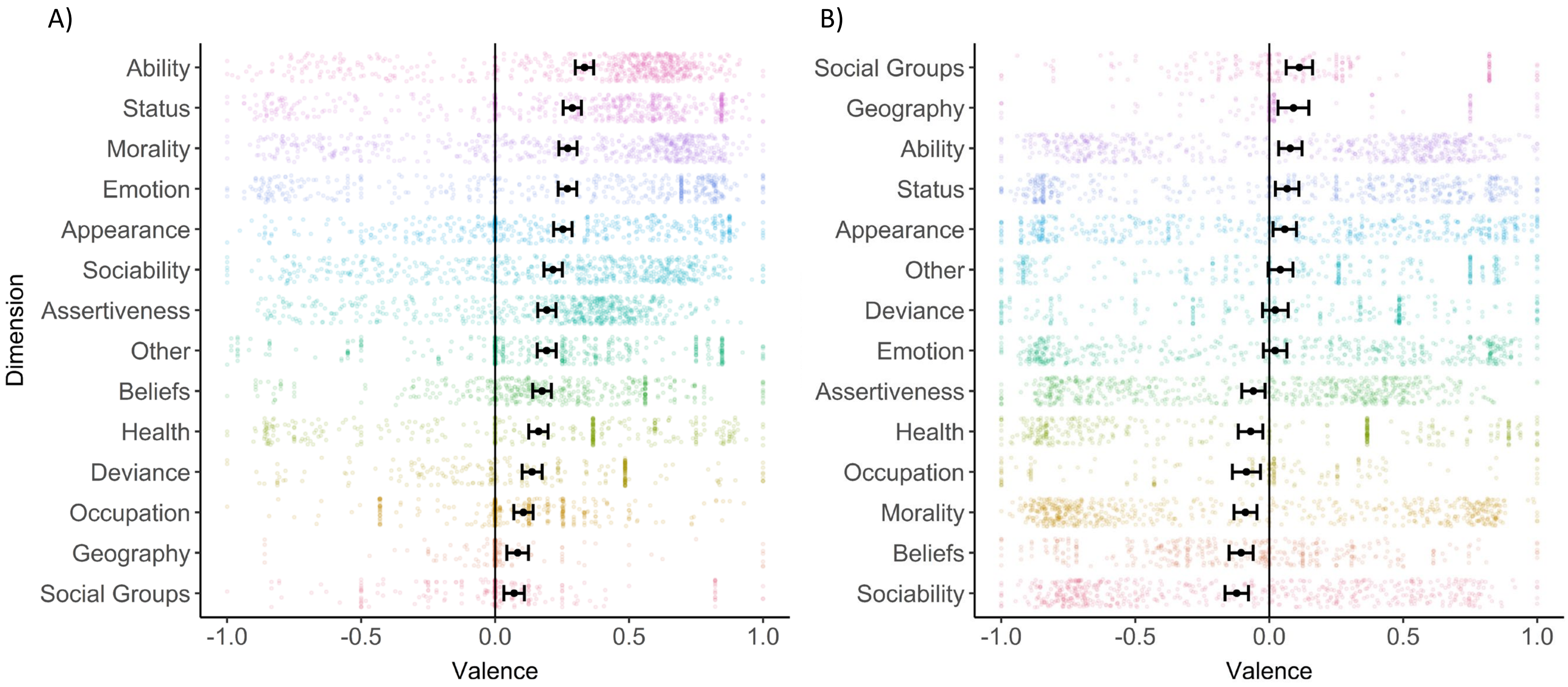

**Fig. 3. Valence for each dimension of the taxonomy.** Valence ranges from -1 (negative) to +1 (positive). Error bars indicate Standard Errors. Panels: a) ChatGPT, b) Llama 3. The Mixtral figure is very similar to the ChatGPT figure and is included in the Supplement.

## Specific Illustrations

In Figure 4, we show examples of stereotype prevalence and direction/valence for specific salient social categories. These results illustrate how understanding both prevalence and direction/valence across multiple dimensions reveals stereotype differences, including higher prevalence of Sociability content for women (vs. men), despite similar direction, or high frequency of content for dimensions beyond Warmth and Competence, such as Deviance stereotypes for “heterosexual” and Emotion stereotypes for “poor” categories. Additional examples are available in the online repository.

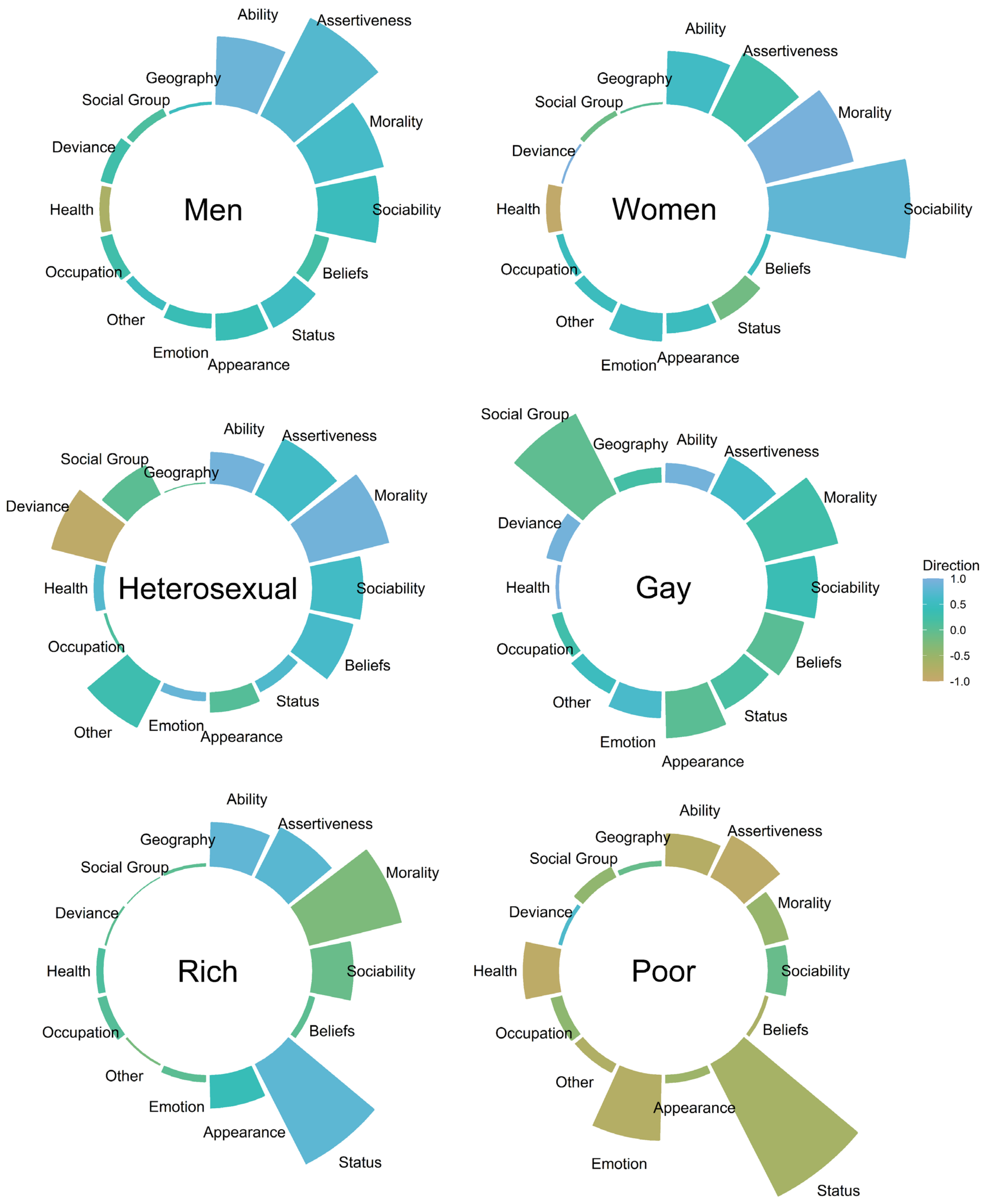

**Fig. 4. Illustration of stereotype profiles for select social categories, ChatGPT**. Bar size indicates prevalence of the dimension. Color indicates direction when available (see Figure 3), otherwise, color indicates valence as a directional proxy.

**Change Over Responses**

For this exploratory analysis we found that there were significant increases (Sociability, Morality, Assertiveness, Emotion, and Other content, *p*s < .007) and decreases (Ability, Status, Beliefs, and Social Groups, *p*s < .027) in content representativeness over the 50 responses for various dimensions. In fact, these patterns (e.g., Ability more common in earlier responses, Warmth facets more common later on) replicate human patterns (*4*). However, the changes were not large enough to meaningfully change the taxonomy prevalence if fewer than 50 responses are considered. Extrapolating from these patterns (with caveats for predicting beyond the data range), we would not expect that moderate increases beyond 50 would be particularly impactful either. Perhaps the exception for change-over-response adjustments would be for Sociability content, which shows the largest change by far, from 7.8% of first responses being about Sociability, increasing to 17.5% of $50^{th}$ responses. See the Supplement for further information.

**Predictive Value**

Finally, we examine whether the full taxonomy improves predictions of LLM and human general valence evaluations of social categories, as compared to using only the valence of the associations as a predictor. As expected (see Table 5), a baseline linear regression model predicting internal general LLM valence about a category from the valence of the responses about the category was significant. However, a regularized model adding all the taxonomy's dictionary-coded dimensions significantly improved predictivity. All dimensions of the taxonomy were retained in at least one LLM, and most were also significant in at least one linear model, in support for both the importance of an extended taxonomy in improving predictions of category representations, and the potential for cross-model variability.

An exploratory analysis using human survey evaluations of general valence provided congruent results, supporting an improvement in predictive power from the expanded taxonomy. A simple model using the valence of the responses to predict general human valence evaluations had $R^2$ = .553 and AIC = 128.12 for ChatGPT, $R^2$ = .277 and AIC = 1360.84 for Llama 3, and $R^2$ = .29 and AIC = 1417.9 for Mixtral (with significant effects for the predictor, $p < .001$). A model adding all the dictionary-coded prevalences had $R^2$ =.620 and AIC =, $\chi2(15) = 6.5$, $p = .01$ for ChatGPT, $R^2$ =.325 and AIC = 1340.39, $\chi2(14) = 22.92$, $p < .001$ for Llama 3, and $R^2$ =.4 and AIC = 1329.4, $\chi2(15) = 51.105$, $p <.001$ for Mixtral.

**Table 5. Prediction of models' internal general valence.**

| Dimension | ChatGPT | Mixtral | Llama 3 |
|---|---|---|---|
| **Ability** | 0.12 | 0.22* | 0.154* |
| **Appearance** | 0.058 | 0.1* | 0.002 |
| **Assertiveness** | 0.015 | 0.041 | 0.115* |
| **Beliefs** | 0.1* | 0.112* | |
| **Deviance** | 0.033 | 0.06* | 0.036* |
| **Emotion** | 0.112* | 0.104* | 0.03 |
| **Family** | | 0.1* | |
| **Geography** | 0.018* | 0.054 | 0.026 |
| **Health** | 0.073 | 0.063 | 0.029 |
| **Morality** | 0.238* | 0.21* | 0.167* |
| **Occupation** | 0.079 | 0.053 | 0.049 |
| **Other** | 0.066* | 0.074* | 0.018 |
| **Sociability** | 0.094 | 0.1* | 0.034 |
| **Social Groups** | 0.013 | 0.092* | 0.037* |
| **Status** | 0.114* | 0.05 | |
| **Baseline $R^2$** | 0.24 | 0.275 | 0.289 |
| **Full $R^2$** | 0.526 | 0.426 | 0.343 |

*Note*. Values are correlations between the internal general valence of a model and the prevalence, direction, or valence (highest selected) for each dimension. For each LLM, if a dimension lacks a value is because the dimension was dropped from the regularized regression. * indicates that the dimension was also significant in the multiple regression model. $R^2$ values compare the baseline regression (only a valence predictor) and the full, regularized model, with all the dimensions.

## Discussion

The present study introduces a nuanced taxonomy of the stereotype content in contemporary LLMs. We prompted ChatGPT 3.5, Llama 3, and Mixtral 8x7b to provide cultural stereotypes about a large number of salient social categories. We then content-coded these associations to understand stereotype associations encoded by the LLMs.

The dimensions of the taxonomy were initially identified via a data-driven cluster analysis of text embeddings of the associations, and largely align with the content of human stereotypes described by the SSCM (*4*). Various dimensions shifted slightly in representativeness between LLMs, analytical approach, and human data, but the general pattern involved very high representativeness of Warmth and Competence facets, followed by Status, Beliefs, Emotions, and Appearance, and then a variety of smaller yet significantly prevalent content, such as Health, Occupations, Deviance, Social group membership, and Geographical stereotypes.

We demonstrate significant comprehensiveness of these dimensions by establishing that they account for ~90% or more of the social category associations. For comparison, the "big two" of Warmth and Competence, recently examined in LLM studies (*17*, *18*), accounted for only about 54-63% of content. Even then, our results suggest differences in representativeness, direction, valence, change over time, and predictivity for facets of Warmth and Competence (Sociability, Morality, Ability, Assertiveness), suggesting that breaking down the "big two" into their theoretical facets (*3*) may be preferrable to characterize stereotype content.

Despite similarity to human models in the representativeness of dimensions, the LLMs had noticeable departures in direction and valence: while human stereotypes tend to be negative, the LLMs showed more positivity across dimensions. This may be the result of reinforcement learning from human feedback and other safeguards put into place after initial model training (i.e., they may not reflect positivity in the training data, *10*, *12*). For Llama 3 this pattern was attenuated, potentially due to the use of a base (vs. chatbot) model, including fewer valence safeguards. Regardless, these findings indicate neutral-to-positive valence for stereotype content across categories and dimensions, on average.

This positivity, however, does not signify a lack of potential for valence-based bias. We find between-category direction/valence differences aligning with expected biased valence evaluations, such as poor people being stereotyped more negatively across most dimensions as compared to rich people (*4*). Additionally, direction and valence differences between dimensions underscores that dimension-specific valence is more informative than the general valence often used in LLM auditing. Some of these differences between dimensions (particularly for direction) align with human patterns such as higher positivity for Competence-related dimensions (Ability, Assertiveness, Status) than Warmth-related dimensions (Sociability, Morality). Future research should further examine these patterns in models with differing levels of safeguards.

In addition, we find differences in prevalence between categories across dimensions, even when the categories are relatively similar on valence. For example, ChatGPT's stereotypes about women focused more on sociability, while men stereotypes focused more on assertiveness; stereotypes about gay people were more about social category membership, while stereotypes about heterosexual people were more about (lack of) deviance (e.g., "average," "normal"). This finding underscores the need to consider both direction/valence and representativeness to more fully capture biased representations (*38*).

Analyses of change over responses suggests that the taxonomy is largely robust to the number of responses requested, with all dimensions remaining statistically significant. However, we did find evidence for dynamic changes across dimensions, which largely align with human patterns. For example, Ability and stereotypes that seem to be more structural (e.g., Status, Beliefs) tend to happen earlier in the list of responses, whereas more interpersonal stereotypes tend to happen later. Earlier responses may also have a bigger impact in AI applications, since shorter text may only retrieve initial stereotypes. Future studies can further examine these temporal dynamics.

Finally, supporting the relevance of the taxonomy in understanding LLMs' stereotypes, the dimensions predicted the LLMs' internal general valence representation of the social categories, above-and-beyond the valence across stereotypical associations. That is, these dimensions provide unique predictive value about the positivity/negativity with which the LLMs (and humans, as suggested by exploratory results) represent social categories. These results support the validity of the taxonomy as representing consequential dimensions, that improve connections between human and LLMs' stereotypes to improve auditing efforts, and between dimension-specific and broader (e.g., general valence-based) representations of social categories.

**Implications**

The taxonomy introduced here provides a more nuanced view of stereotype associations in LLMs. While most previous research, auditing, and debiasing efforts have tended to focus on general valence patterns, our paper suggests that understanding the content of LLMs' associations with social categories requires a much wider set of content dimensions. These same dimensions have been found to describe open-ended stereotypes in humans (*4*), as well as face impressions and other person perceptions (*39*), in support of their relevance.

In previous research, these dimensions have been shown to predict prejudice towards and decision making about social targets. Warmth and Competence facets are well-established predictors of outcomes ranging from hiring and performance evaluations to negative interpersonal behaviors (*30*, *32*). Moreover, the rest of the taxonomy predicts scenario-based decision-making outcomes such as which social categories to prioritize for policies guaranteeing access to healthcare (Health dimension), protection from hiring discrimination (Social groups, Geography), or protection from discrimination in facial recognition technologies (Appearance, *4*, *39*). Understanding how these dimensions are reflected in LLMs can expand the ways in which we measure stereotypes relevant to these outcomes (e.g., over time, across languages), with implications for social psychological theory and interventions (*40*–*43*). However, such inferences from LLM to human cognition must carefully consider training data (transparency and biases), fine-tuned safeguards that may distort cultural patterns, and the potential for LLMs reflecting novel or distinct stereotypes due to how they synthesize and process information in non-transparent ways (e.g., *44*).

More directly relevant to LLM development and use, a deeper understanding of the nuances of stereotypes can help prevent bias percolating through auditing and debiasing approaches focused on general indicators. Developing benchmarks and debiasing procedures that address higher-dimensional stereotypes and multiple properties, including representativeness, direction, and valence, will provide a more accurate picture of fairness in LLMs and responsible applications. For example, auditing efforts should pay attention not just to general positivity in categories' representations in LLMs, but valence across many relevant stereotype dimensions, and our taxonomy provides an initial set of content to evaluate. Similarly, auditing should go beyond valence/direction for these dimensions, to also measure prevalence, providing fuller profiles of the representation of social categories (*38*). This would identify, for example, men and women showing similar valence across dimensions, yet being described more often based on some dimensions over others (e.g., women associated more with Warmth and emotions, and men associated more with Competence). These steps may reduce harmful exposure to stereotypes for stigmatized groups, reductions in the perpetuation of stereotypes via AI, and improved human-computer interaction, among others.

**Limitations and Future Directions**

The current research is not without limitations. First, our results are US- and English-centric, based on the training data of the LLMs selected biasing heavily towards these data. However, initial cross-cultural research with human participants showed fair stability of the SSCM taxonomy (*4*). Second, the LLMs used show a significant lack of transparency regarding training data and implemented safeguards. As such, our ability to connect LLM representations to cultural representations is limited. Third, we restrict our results to three models in a growing field of LLMs. However, the striking consistency between these independent models, and human data, suggests that this taxonomy may be robust, with variability across specific properties (e.g., valence) to be studied in future research. Fourth, our prompts focused on cultural associations rather than "personal" ones. However, LLMs may not make significant distinctions in their representations of concepts based on whether instructions request cultural vs. an artificial "personal" representation. In fact, preliminary analyses suggest that at least some stereotype properties are invariant to these prompt variations (*38*). Nonetheless, more research is needed. Here, we focused on maximizing convergence with human data, which has most often asked about cultural stereotypes, in part to minimize socially-desirable responding (c.f., chatbot safeguards), but also because cultural stereotypes have been shown to predict the myriad of outcomes discussed previously. A related future direction should explore the taxonomy in less explicit prompts that may elicit more negative stereotypes (e.g., *45*). Fifth, open-ended data are difficult to analyze. In line with the SSCM, we employed a multimethod approach to examine robustness to analytical methods and find minimal differences (text embeddings clustering and dictionary analyses). However, future studies may improve coding methods to potentially extract more information from the models.

Additional future directions include expanding the taxonomy to other multimodal and intersectional targets (*46*, *47*) and developing auditing and debiasing methods incorporating the taxonomy. Finally, we note that the current taxonomy includes dimensions that could always be further broken down or combined. This taxonomy aims to balance nuance with a manageable number of dimensions, but for those interested, additional subdimensions exist in the dictionaries (or instruments could be developed). For example, Appearance may be broken down into Attractiveness, Clothing, Body Properties (see online repository, and previous research, *48*). On the other hand, dimensions may be further combined for more parsimonious solutions, depending on the need for nuance (e.g., creating Warmth and Competence dimensions instead of using their facets).

A complete understanding of the biases encoded into increasingly influential AI technologies requires acknowledging the multidimensional nature of stereotyping. The LLM stereotype taxonomy we identified and characterized largely aligns with human models, such as the SSCM, while showing differences in valence and direction of stereotypes. As LLMs continue to be developed and deployed, our findings suggest that auditing and debiasing efforts should attend to the complexities of stereotypes, in an effort to minimize their harmful consequences.

## Materials and Methods

The study uses data from LLMs to conduct a quantitative content analysis via text embeddings and dictionary coding, revealing information about the coverage, prevalence, direction, and valence of the identified taxonomy of LLM stereotypes.

All data and materials available in the online repository: https://osf.io/bdu6g/?view_only=733e364b751a4736af2c3c897cf476c8

**Stimuli**

We used a list of 1,366 different terms referring to 87 salient social categories in the U.S. For example, terms such as “wealthy,” or “millionaire” were stimuli used to represent the “rich” social category. These terms have been validated and used successfully in previous LLM studies to elicit stereotype content (*38*). See Table 6 for categories and the Supplement for a full list of terms.

**Table 6. List of social categories used as stimuli and labels for related terms.**

| | | | | |
|---|---|---|---|---|
| Home-schooled | Lower-class | Ivy-leaguers | Mentally Handicapped | Preps |
| Teenagers | Athletes | Investors | Atheists | Educated |
| Accountants | Black | Geeks | Elderly | American |
| Democrats | Gay | Women | Christians | Students |
| White-collar | Mexican | Disabled | Drug addicts | Heterosexual |
| Gamers | Hippies | Blue-collar | Blind | Lawyers |
| Celebrities | Working-class | Parents | Doctors | Engineers |
| Libertarians | Asian | Scientists | Middle Eastern | Upper-class |
| White | Muslim | Jewish | Hindus | Independents |
| Nurses | Hipsters | Poor | Goths | Welfare recipients |
| Children | CEOs | Jocks | Indian | Adults |
| Vegans | Immigrants | Men | Buddhists | Rich |
| Crossdressers | Republicans | Middle-class | Obese | Nerds |
| Politicians | Hackers | Criminals | Germans | Teachers |
| Catholics | Liberals | Bisexual | Religious | Conservatives |
| Hispanic | Transgender | Homeless | Unemployed | Rednecks |
| Young | Artists | Lesbians | Musicians | Sex workers |
| Bankers | Native American | | | |

## Language Models

We primarily focus on ChatGPT for analyses but provide results from two independent models, Mixtral 8x7B, and Llama 3 for coverage, prevalence, direction, valence, and predictive analyses for insights across current LLMs.

### *ChatGPT*

We use GPT 3.5 turbo as implemented in freely-available versions of ChatGPT as of July 30th, 2024 (*10*). The ChatGPT model was trained on vast amounts of data, including the Common Crawl (a large scraping of internet webpages), books, Reddit, and Wikipedia (*11*), as well as human feedback in reinforcement learning (*12*), and potentially others. We used the Python OpenAI API to access the model.

### *Mixtral*

We use Mixtral 8x7B (with instruct fine-tuning for chatbot functionality), accessed via Python Transformers and HuggingFace. Mixtral is a "sparse mixture of experts" model (*37*, *49*). Unlike the ChatGPT model used here, Mixtral has open weights, providing some additional level of transparency. Mixtral also has either similar or superior performance to ChatGPT 3.5 on various benchmarks (*37*). As with ChatGPT, the training data for the model is not transparently disclosed by the developers, but is described as being extracted "from the open Web." (*49*)

### *Llama 3*

We use Llama 3 8B (*50*) as another open-source LLM, accessed via Python HuggingFace. Llama 3 uses a decoder transformer architecture and achieves similar or superior performance to other similarly-sized open-source LLMs. Unlike for the previous models, the Llama 3 model used here is the base model (i.e., not fine-tuned to behave as a chatbot). Using a base model allowed us to reduce the influence of potential safeguards introduced via chatbot fine-tuning. However, prompts about internal valence (see below) produced low-quality data (e.g., primarily, failure to provide a response) for the base model, so we relied on the fine-tuned "instruct" model for this variable.

Available information about the Llama 3's training data includes that it consisted of over 15T tokens from public data, over 5% non-English data, involving some filtering (e.g., NSFW filters), and other quality checks (*50*).

**Prompts**

All prompts use language from human research requesting the LLMs to provide characteristics that most Americans associate with social categories. In survey research, this approach allows for measurement of (perceptions of) "cultural stereotypes" shared across people, and helps reduce social desirability concerns of providing personal stereotypes (*6*). These cultural stereotypes are nonetheless highly predictive of bias and discrimination intent (*4*, *5*, *30*). Prompt language referring to cultural stereotypes may also reflect the aggregate (vs. individualized) nature of the associations LLMs learn.

As a "system prompt", we entered: "You will be given information about a series of target groups of people. Answer how most people in American society would respond."

Following the "system prompt," for each stimulus term, we retrieved their stereotype associations by prompting: "List 50 characteristics that you believe most Americans would think describe [*target*]. Use single words". We prompted the LLM for 50 characteristics to obtain an expansive representation of the top stereotypical associations. Human research usually asks for far fewer associations (e.g., one to ten, *4*), and in general, the LLMs often provided fewer than 50 responses, potentially when no additional associations were available. To further examine the role of number of responses on the taxonomy, we also present a change-over-response analysis of content in the Supplement (Changes in prevalence tended to be small, not significantly impacting the ranking of most dimensions, with the exception of Sociability associations which became more common in later responses and may thus be less prevalent if fewer responses are requested).

To retrieve the model's internal general valence evaluations of the categories, we prompted: "In general, how does American society view people who are [*target*]? Use a scale ranging from 1) Very negatively to 5) Very positively. Do not provide an explanation, only a single-number response using the scale."

To obtain the most deterministic results, we set the temperature (a parameter that manipulates the randomness of the models' output) to 0, when possible. Repeated associations within a response were removed if they occurred. Some output included warnings about bias, which were removed. In addition to warnings, the models failed to return lists for terms it indicated are "not commonly used or understood in American society" (e.g., "mahanaya" for ChatGPT). For all 87 categories, except the "Black" category in ChatGPT (which returned only warnings and was thus removed from subsequent analyses), we successfully retrieved the requested output for at least one term. Because the Llama 3 base model is trained for sentence completion rather than chat (*50*), we slightly modified the prompt for sentence-completion rather question answering (see Supplement for these variations).

**Human Data**

For obtaining human stereotype dimension prevalence, direction, and valence (as descriptive baselines; from open-ended items requesting the characteristics that come to mind to describe the categories), as well as general valence evaluations of social categories (how participants evaluated the categories in general, on a scale from 1 – Very Negatively to 5 – Very positively), we use two published studies that contain parallel data on the same social categories (*4*, studies 3 and 3R, N = 797, including 4,782 category ratings). All human measures used the same format as for the LLMs.

**Statistical Analysis**

We preprocessed all responses for the stereotype associations by transforming words from plural to singular, removing capitalization, and replacing dashes with spaces.

***Obtaining Text Embeddings***

To run the initial cluster analysis, we first obtained the text embeddings for the LLMs' responses. Text embeddings are numerical vector representations of each stereotype association, encoding information about the semantic relations between words.

Here, we use the embedding model SBERT (*51*, *52*). SBERT embeddings have fewer dimensions than those underlying the LLMs used here (making them more suitable for cluster analysis), are openly available (unlike ChatGPTs'), and have shown validity in previous cluster analyses of social perceptions (*4*, *39*). In addition, by using a different model to code the responses, our coding is independent from the internal representations of the LLMs used here, avoiding potential "double dipping" on a model's bias. However, we note that because the cluster analysis uses only the responses, without connecting them to the categories they were provided for, it captures the semantic structure of the responses, not biases based on the category-response association.

***Cluster Analysis***

With the embeddings, we computed a (dis)similarity matrix using pairwise cosine similarities between all of the unique response embeddings (N = 5,871). For example, words such as *fit* and *healthy* received higher cosine similarity scores than pairs such as *fit* and *black* *hair*. We computed a k-means clustering solution from the dissimilarity matrix. To select an appropriate number of clusters (*k*), we used the R package NBclust (*53*), which runs multiple metrics of cluster fit. NBClust suggested $k = 2$ (6 metrics) and $k = 59$ (5 metrics) most frequently from the *k*s tested (2 through 60). Given a preference for higher-cluster solutions to capture more diversity of content, we set $k = 59$. Then, to facilitate labeling by the researchers, we obtained, for each cluster, the responses that were most semantically similar to their centroid (i.e., those that were most prototypical of the cluster). Finally, we used the k-means results to code all responses and compute cluster analyses.

***Dictionary Coding***

Dictionary approaches allow measurement of coverage, direction, and valence, and provide an independent coding approach from the cluster analysis. Dictionaries are lists of words coding for content referring to the dimensions identified in the cluster analysis, and which can be matched to the LLM responses. The dictionaries used here have been validated (*48*) and used successfully in previous studies of stereotypes in LLMs (*38*). The dictionaries include over 15,700 terms coded into over 14 stereotype dimensions, obtained via automated lexical expansion of theoretical seed words using WordNet (see Supplement for additional information on the structure of the dictionaries and validation data). Human stereotypes coded via the dictionaries predict relevant outcomes, including decision-making and general prejudice (*4*).

The dictionaries code separately for representativeness and direction. For representativeness, a response that is present in a dictionary receives a score of 1 for the corresponding dimension (0 if absent). For example, if “amicable” or “unfriendly” are stereotypes, they would receive a score of 1 for Sociability and 0 for other dictionaries, since they are words in the Sociability dictionary, but not in others. For other examples, “weird” and "average" code into the Deviance dictionary; "smart" and "unintelligent" into the Ability dictionary. Some words may fall into multiple dimensions/dictionaries. LLM responses that could not be coded into any of the dictionaries were marked as "No match," and this value is used to calculate coverage (i.e., 1 – percentage of no-matches).

Dictionaries also coded responses on direction per dimension. Each response was coded for direction as -1 (low) to 1 (high), or as missing data if the response was not about the dimension (i.e., if it received a score of 0 for representativeness). Thus, responses such as "friendly" or "amicable" received direction scores of 1 for Sociability, while "unfriendly" received a score of -1, and were coded as missing values for other dimensions. Valence for each response was coded similarly but with a continuous variable ranging from – 1 (negative) to 1 (positive).

Because the LLMs were prompted to provide 50 responses per category term, we averaged scores across responses.

***Regression Models***

Given the large number of observations, we had power > 90% for all tests, using a small-to-medium effect size ($r$ = .2), as indicated by the R package *simr* (*54*). The exception were tests using human data as an outcome, due to only having available data for the 87 overarching categories, resulting in 80% power for a medium effect size ($r$ = .3).

In general, we use mixed regression models with category as a random factor (to account for non-independence), and each term/exemplar as an observation. However, for models with level 2 outcomes (e.g., when predicting human general evaluations, which had data for one term per category), we aggregate the data at the level of the 87 overarching categories (i.e. collapsing across their terms) and run linear regressions with robust standard errors (*55*).

For predicting internal valence, we first model only general valence across responses (as coded via dictionaries) predicting LLM general valence evaluation of the category (as measured by the item “In general, how does American society view people who are [*target*]? Use a scale ranging from 1) Very negatively to 5) Very positively.”).

In a second step we use regularized models (to further account for number of predictors and overfitting), using the scores for all dimensions on representativeness, direction, and valence as predictors (for dimensions where direction was available, we used it, otherwise we used valence). We used the glmnet R package (*56*) for regularized regression and selected the lambda that minimized mean cross-validated error for each model. We compare the $R^2$s for the general valence-only model with the $R^2$s from the full regularized model as our main test, but also present dimension-specific correlations with general valence.

For change-over-responses, we use mixed models with term and category as random intercept, dimension name and response order, ranging from 1 (the first response provided) to 50 (last response) for each term as interacting predictors, and prevalence of each dimension as outcome.

For all models, we present additional details and statistics in the Supplement.